\documentclass[12pt]{iopart}
\usepackage{graphicx}

\begin{document}

\title{Loss-induced lasing: new findings in laser theory?}

\author{Stefano Longhi and Giuseppe Della Valle}

\address{Dipartimento di Fisica, Politecnico di Milano and Istituto di Fotonica e Nanotecnologie del Consiglio Nazionale delle Ricerche, Piazza L. da Vinci
32, I-20133 Milano, Italy}
\ead{longhi@fisi.polimi.it}
\begin{abstract}
In a recent work, using a coupled microresonator system with tailored gain and loss parameters B. Peng et al. [{\it Science} {\bf 346}, 328 (2014)] have experimentally reported on an apparently counterintuitive effect in laser theory,  namely the possibility to enhance lasing by increasing loss in the system. The observed phenomenon was related to the existence of an exceptional point in the system and was presented somehow as an unexpected and novel effect, especially by some reporters and scientific blogs. In this communication it is pointed out that the phenomenon of loss-induced lasing does not come as a surprise in known laser theory and that it is not necessarily related to the physics of exceptional points. Loss-induced lasing is basically the lasing mechanism that occurs in loss-coupled distributed feedback lasers. This mechanism dates back to the 1970's, has a simple physical explanation and does not rely on the physics of exceptional points. 
\end{abstract}



\maketitle

Lasing is based on the interplay between optical
gain and optical feedback in a cavity.  Once the gain provided  
through external pumping overcomes
the loss in the cavity, lasing sets in
for the spatial and frequency cavity mode with the lowest threshold \cite{r1,r2}. Thus, loss is a natural adversary
of lasing, at least in common lasers. However, this is not always the case when losses (or gain) are not uniformly distributed in the medium. Using a pair of coupled optical micro-resonators with tailored gain and loss,  in a recent experiment \cite{r3}  Peng et al. 
 reported a violation of such a simple picture, demonstrating that 
 by steering the parameters of the system to the vicinity of an exceptional point (EP) lasing can be enhanced when the total loss of
the system is increased. More precisely, the authors found that an increase of total loss in the system first  annihilates an existing Raman laser, however beyond a critical value  lasing recovers despite the increasing loss. Other unusual effects in the laser behavior, related to the appearance of EPs, have been emphasized in other recent papers as well, such as the anomalous dependence of  laser output power on pumping level and pump-induced lasing death \cite{r4,r5,r6,r7}. Such experimental findings, and especially the fact that losses can enhance (rather than suppress) lasing, are attracting considerable resonance since they apparently seem to violate some basic picture of lasing that we have learned in textbooks, as some reporters and scientific blogs wrote commenting on such experimental findings  \cite{r8,r9,r10}. While the above mentioned effects are surely uncommon and recent experiments have beautifully used integrated photonic structures to visualize the physics of non-Hermitian and $\mathcal{PT}$-symmetric systems, as a matter of fact loss-induced lasing is not a new nor a surprising result in laser physics, even at textbooks level. When loss or gain in the system are not uniformly distributed, 
 the lasing mode can avoid to occupy the regions with largest loss, and an increase of total (average) losses in the system could paradoxically, in some cases, yield a reduction of the {\it effective} losses experienced by the mode, i.e. an increase of total losses could enhance (rather than prevent) lasing.  Such a behavior
is not new  in laser physics. It can be observed without resorting to the properties of EPs and it is at work in a kind of technologically important lasers like semiconductor lasers:
 loss-coupled distributed-feedback (LC-DFB) laser (see, e.g., \cite{r11,r11bis,r12,r13,r14} and references therein). LC-DFB lasers were earlier developed for semiconductor lasers as a mean for single frequency operation robust against spatial hole burning, contrary to index-coupled DFB lasers with a uniform index grating that tend to emit into two modes far apart from the Bragg frequency. In LC-DFB lasers losses play the fundamental role of providing optical feedback, thus they are helpful (rather than detrimental) for lasing.
 
 \par 
 It is the aim of this short communication to remark that loss-induced lasing is not a new nor a surprising effect in known laser theory and that it does not necessarily require a system to operate near an exceptional point. In particular, we show that in a LC-DFB structure laser threshold can astonishingly {\it decrease} as the average loss in the system {\it increases}: i.e. lasing can benefit from an increase (rather than a decrease) of total losses in the system like in Ref.\cite{r3}. This is actually nothing magic and a simple physical explanation can be given, which does not rely on concepts like EPs taken from non-Hermitian physics.  
 
 \begin{figure}
\includegraphics[scale=0.4]{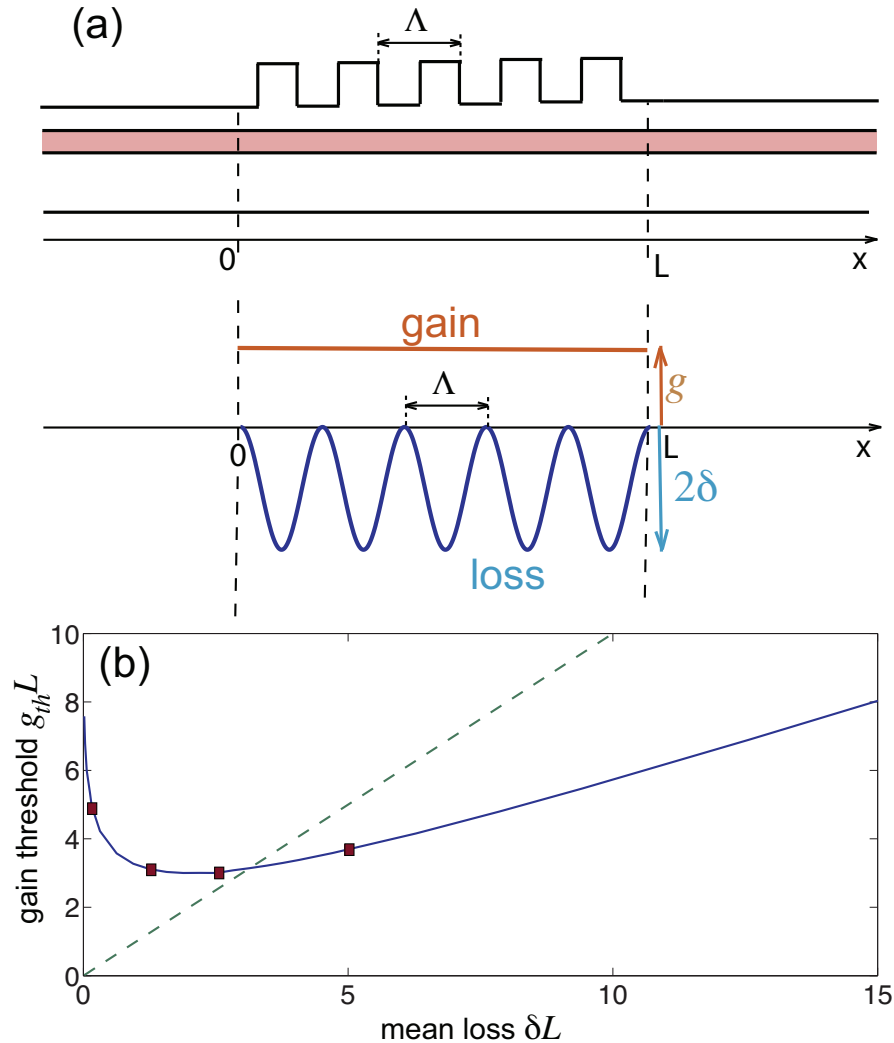}
\caption{(Color online) (a) Schematic of a DFB structure with uniform gain and with a loss grating. (b) Behavior of the laser threshold $g_{th}L$ versus the mean loss of the grating $\delta L$ (solid curve). Above (below) the dashed curve the lasing threshold is higher (smaller) than the mean loss. $g_{th}=\delta$ occurs at $\delta L= \pi$. For the values of $\delta L$ corresponding to the squares on the solid curve, the detailed behavior of the  spectral transmittance in the complex frequency plane is depicted in Fig.2.}
\end{figure}

 \begin{figure}
\includegraphics[scale=0.25]{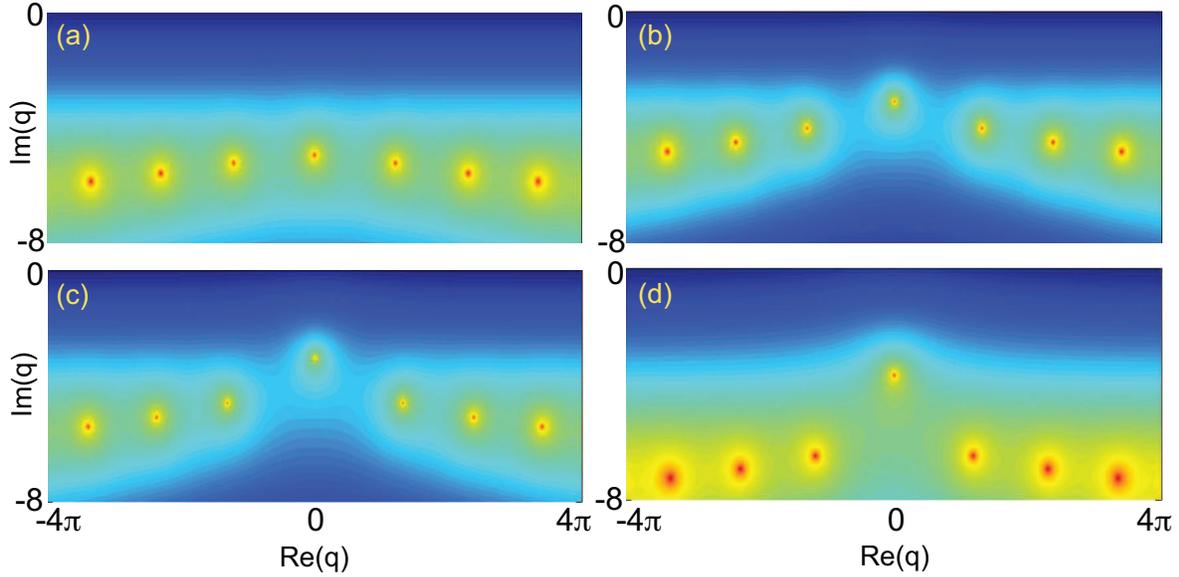}
\caption{(Color online) Snapshots of $|t (q)|$ in the complex $q$ plane (pseudocolor maps in log scale) for increasing values of $\delta L$ [square points in Fig.1(b)]: (a) $\delta L =0.05 \pi$, (b) 
$\delta L =0.4 \pi$, (c) $\delta L =0.8 \pi$, and (d) $\delta L =1.6 \pi$.}
\end{figure}

 Let us consider one-dimensional wave propagation in a dielectric waveguide, and let us assume that some spatially uniform optical gain $g$ per unit length, spectrally flat  at around a frequency $\omega_0$, is applied by some external pumping in the waveguide region $0<x<L$  of length $L$ [Fig.1(a)].  Such a system clearly behaves like an optical amplifier, however lasing can not occur because of the absence of any optical feedback (infinite output coupling losses). Optical feedback can be introduced by adding some refractive index discontinuities (e.g. Fresnel reflections from the facets of a semiconductor/air interface), or by distributed optical feedback via a periodic index grating. However, the same goal can be achieved by means of a {\it loss grating}: in this case loss-induced lasing can be realized. Let us introduce in the waveguide region $0<x<L$ a loss grating with spatial period $\Lambda= \pi / k_0= \pi  c_0 / (n_0 \omega_0)  $ and amplitude $\delta$, yielding a spatially-dependent loss rate per unit length $\gamma(x)=\delta[1-\cos(2 \pi x / \Lambda)]$, where $n_0$ is the mode index of the waveguide and $c_0$ is the speed of light in vacuum. Clearly the grating introduces spatially-dependent losses in the system that vary periodically between zero and $2 \delta$ and with a spatial average loss rate $\delta$.  Despite the grating introduces losses into the dielectric waveguide, it provides some distributed optical feedback, thus allowing lasing at the Bragg frequency $\omega_0$.  {\it Losses can be thus turned into effective gain for lasing}. The laser threshold $gL=g_{th}L$ turns out to be a function of $\delta L$ solely and its behavior is shown in Fig.1(b).  In the figure the behavior of the {\it spatial average} losses in the system $ \delta L$ is also depicted by the dashed curve.  The threshold curve can be calculated by standard coupled-mode theory \cite{r11,r11bis,r12}. The electric field $E(x,t)$ in the one-dimensional guide structure can be written as a superposition of counter-propagating waves
 \begin{equation}
 E(x,t)=\psi_1(x,t) \exp(-i \omega_0t-ik_0 x)+\psi_2(x,t) \exp(-i \omega_0 t +i k_0 x) +c.c. \;\;\;\;
 \end{equation}
 where the slowly-varying amplitudes $\psi_{1,2}$ of counter-propagating waves satisfy coupled-mode equations
 \begin{eqnarray}
 \frac{\partial \psi_1}{\partial x}+\frac{1}{v_g} \frac{\partial \psi_1}{\partial t} & = & (g-\delta) \psi_1+(\delta / 2) \psi_2 \\
- \frac{\partial \psi_2}{\partial x}+\frac{1}{v_g} \frac{\partial \psi_2}{\partial t} & = & (g-\delta) \psi_2+(\delta / 2) \psi_1
 \end{eqnarray}
 where $v_g \simeq c_0/n_0$ is the group velocity. Equations (2) and (3) hold in the grating region, i.e. for  $0<x<L$, whereas outside the grating region ($x<0$ and $x>L$) the right hand sides of the equations should be replaced by zero. Looking at a solution to Eqs.(2) and (3) of the form $\psi_{1,2}(x,t)=u_{1,2}(x) \exp(-i \Omega t)$, where $\Omega$ is a (generally complex) frequency detuning from the Bragg reference frequency $\omega_0$ , the amplitudes $u_{1,2}$ at $x=0$ and $x=L$ are related by the linear relation
 \begin{equation}
 \left(
 \begin{array}{c}
 u_1(L) \\
 u_2(L)
\end{array} 
 \right)= \mathcal{M}(\Omega) \left(
 \begin{array}{c}
 u_1(0) \\
 u_2(0)
\end{array} 
 \right)
 \end{equation}
  where $\mathcal{M}(\Omega)$ is the $2 \times 2$ transfer matrix. The laser threshold $g=g_{th}$ can be obtained by  looking at the poles of the transmission function $t(\Omega)=1 / \mathcal{M}_{22}(\Omega)$  in the lower half part of the complex $\Omega$ plane (${\rm Im} (\Omega) <0$) by taking $g=0$. Such poles  correspond to resonance modes of the passive structure (see for instance \cite{Longhi}) and are found as the roots  of the transcendental equation
\begin{equation}
\cos (\theta L)+i \frac{\sigma}{\theta} \sin (\theta L)=0
\end{equation} 
where we have set 
\begin{equation}
\sigma=-i \delta-n_0 \Omega / c_0 \;\;, \;\;\; \theta=\sqrt{\sigma^2+\delta^2/4}.
\end{equation}
 The real and imaginary parts of each pole $\Omega$ determine the frequency detuning from Bragg frequency and decay rate of the resonance modes of the passive structure. The introduction of a uniform gain $gL$ just shifts the poles upward toward the real axis, and lasing is obtained when a pole crosses the real axis ${\rm Im}(\Omega)=0$; the lasing mode is thus the one closer to the real axis ${\rm Im}(\Omega)=0$ when $g=0$. After introduction of the normalized frequency detuning $q=(n_0 L / c_0) \Omega $, the imaginary part (in absolute value) of $q$ of the most unstable mode corresponds to the lasing threshold $g_{th}L$ of the resonance mode once the uniform gain $g$ is applied. The most unstable mode always occurs at ${\rm Re}(\Omega)=0$, i.e. at the Bragg frequency. This is shown in Fig.2, which depicts contour-plots  of $| t (\Omega)|$ at a few increasing values of the average loss parameter $\delta L$. Note that, as $\delta L \rightarrow 0$, the lasing threshold goes to infinity [see Fig.1(b)]: this is because in the absence of the loss grating there is not optical feedback in the system and laser can not start. As $ \delta L$ (i.e the mean loss) increases, lasing is possible and, contrary to what happens in a conventional laser system, the laser threshold {\it decreases} as the mean loss  $\delta L$ of the system is increased. The reason thereof is that, while the average losses in the system increase, the optical feedback increases as well thus effectively reducing the {\it output} coupling losses. At $\delta L \simeq 2$ the laser threshold shows a minimum, and as losses are further increased the laser threshold increases. At $\delta L = \pi $ the gain and mean losses equal each other; in this conditon the intensity distribution along the structure is uniform, a feature that was noticed and exploited in earlies studies on LC-DFB lasers for stable single-frequency operation against spatial hole burning \cite{r11}. For $\delta L > \pi$, i.e. above the crossing point of the two curves in Fig.1(b), the gain at laser threshold gets {\it lower} than the average losses in the system, and $g_{th} \simeq \delta /2$ as $\delta L \gg 1$. In a practical device the realization of a pure gain grating is challenging, and nowadays DFB lasers are generally based on index gratings with phase shifts \cite{r11}. Nevertheless such a simple example, that is taken from basic theory of DFB optical structures, clearly indicates that loss-induced lasing is not a new nor a surprising effect in laser physics. The example points out that EPs are not necessarily involved here for the observation of loss-induced lasing: indeed lasing arises here, as in the most common laser systems, by a 
 {\it simple} pole that crosses the real axis, without any coalescence of poles (see Fig.2). Hence no exceptional points arise here.  A simpler and rather general explanation of loss-induced lasing can be given from a different point of view: in any laser system the gain at threshold equals the mode losses, and {\it not} the average cavity losses. The two losses are general distinct because of the inhomogeneous distributions of field and losses. In most conventional laser systems the mode losses increase with the average losses, however whenever the losses affect the mode field distribution this trend can be modified and eventually reversed, such as in the 
  LC-DFB structures where losses are beneficial because they introduce optical feedback.
 
 \par For the sake of completeness, it is worth mentioning that a scenario similar to the one reported in Ref. \cite{r3}, i.e. laser death and subsequent revival as total losses in the system are increased that arises from pole coalescence, could be naturally observed in the LC-DFB system when two low-reflectance mirrors of reflectance $R$ are placed at $x=0,L$. Such a reflectance can arise, for example, from Fresnel reflection at the semiconductor facets. In this case, in the absence of the loss grating ($\delta=0$) laser oscillation can arise for sufficiently high gain $gL$ because of the Fabry-Perot cavity formed by the two mirrors at $x=0$ and $x=L$. As $\delta L$ is increased from zero, the overall losses in the system are increased, a feature which does not favor lasing;  however additional feedback is induced by the loss grating, which can add in phase or in opposite phase with the feedback provided by the mirrors. In the former case the loss-induced feedback enhances lasing in the Fabry-Perot cavity mode at Bragg frequency, whereas in the latter case it counteracts it.
  The interplay among the above mentioned effects determines the lasing threshold, which can be determined by coupled-mode theory considering the modified form of the spectral transmission function $t(\Omega)$ taking into account the effect of additional mirrors, i.e. of the Fabry-Perot cavity. Indicating by $r=\sqrt{R} \exp(i \phi)$ the reflection coefficient of the two mirrors, Eq.(5) is replaced by the following one
 \begin{eqnarray}
 \left[ 1-R \exp(2 i \phi+2 i k_0L) \right]  \cos (\theta L) + \\
 \left[ i \frac{\sigma}{\theta} \left( 1+R \exp(2 i \phi+ 2 i k_0L) \right)  -\frac{\delta}{\theta} \sqrt{R}  (1+ \exp(2ik_0L)) \exp(i \phi)  \right] \sin (\theta L)=0. \nonumber 
 \end{eqnarray}
 For $\delta L=0$, the laser threshold of the Fabry-Perot longitudinal modes is the same for all modes and given by the usual relation $g_{th}L=-(1/2) {\rm ln}(R)$; see Fig.3(a). To show the phenomenon of loss-induced suppression and revival of lasing, let us consider the case $k_0L= n \pi$ ($n$ integer number), i.e. the cavity length $L$ is an integer multiple than the grating semi-period $\Lambda$, and $\phi= \pi$. This case corresponds to distributed feedback for the Bragg mode induced by the grating which is in anti-phase as compared to the optical feedback provided by the Fabry-Perot cavity. 
 \begin{figure}
\includegraphics[scale=0.25]{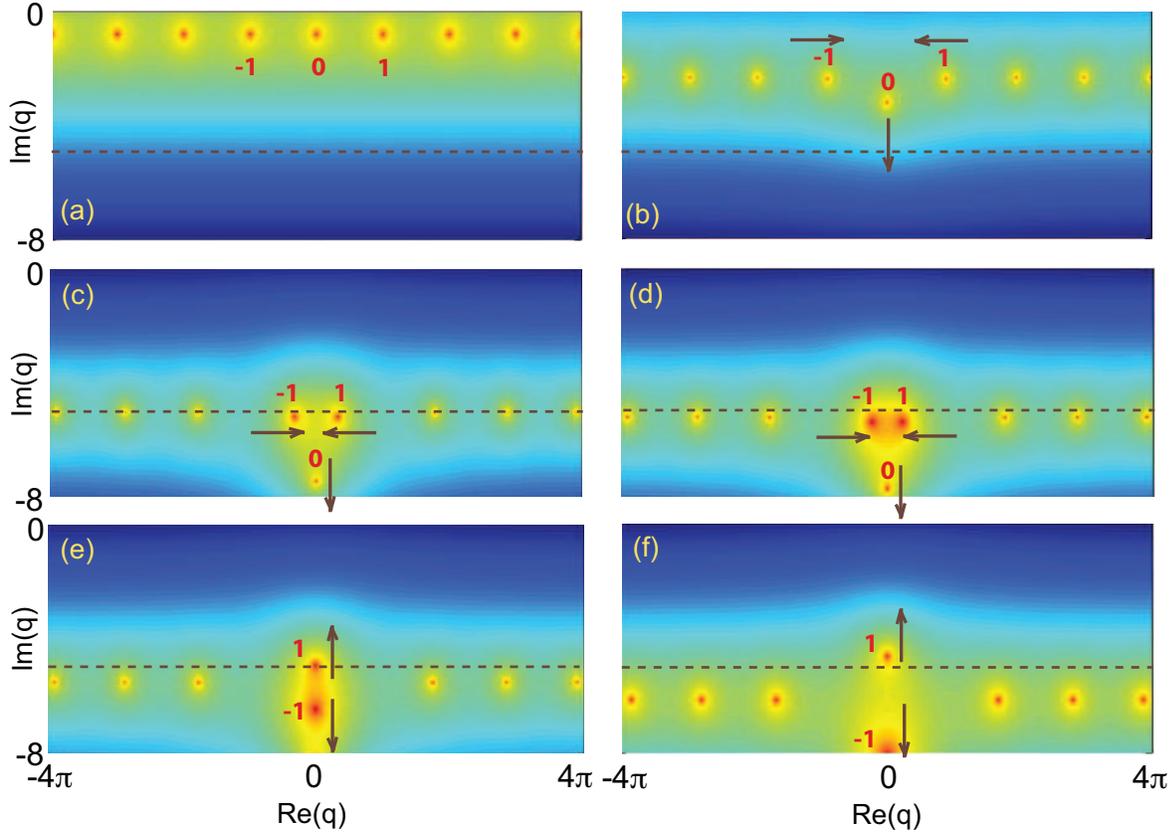}
\caption{(Color online) Same as Fig.2, but for a LC-DFB grating with two additional mirrors at $x=0$ and $x=L$ (reflection $r=\sqrt{R} \exp(i \phi)$). 
 (a) $\delta L =0$ (Fabry-Perot cavity without the loss grating), (b) 
$\delta L =0.5 \pi$, (c) $\delta L =1.35 \pi$, (d) $\delta L =1.4 \pi$, (e) $ \delta L=1.5 \pi$, and (f) $\delta L= 1.7 \pi$. The other parameter values are $R=0.2$, $ \phi= \pi$, and $k_0L$ integer multiple than $\pi$. Labels $0$ and $\pm 1$ highlight the central (Bragg) and two lateral longitudinal modes of the Fabry-Perot cavity. The arrows in the panels show the motion in the complex plane of the poles associated to such three modes  as $\delta L$ is increased. In panels (e) and (f) the mode 0 has an imaginary part smaller than $-8$ and it is thus no more visible. The horizontal dashed lines depict the level of uniform gain ($gL=5$). Poles above (below) the dashed curve are above (below) laser threshold.}
\end{figure}
Figure 3 shows, as an example, a few
 snapshots of the spectral transmission $|t(q)|$ in the complex $q$ plane for increasing values of $\delta L$.  The horizontal dashed line in the figures indicates the value of the uniform gain parameter $gL$, which is chosen in the example equal to $gL=5$. Any pole of the transmission function that lie above the dashed curve is a lasing mode, whereas the poles below the dashed curve are resonance states, i.e. they are below threshold for lasing. As it can be seen from an inspection of the figure, at $\delta L=0$ all the longitudinal  modes of the Fabry-Perot cavity have the same threshold and can lase [Fig.3(a)]. At low values of $\delta L$ the system is still lasing [Fig.3(b)], and the role of the grating is to increase the threshold of all modes and to make the central (Bragg) mode, labelled by 0 in the figure, the highest-threshold mode. Such a behavior stems from the fact that the Fabry-Perot modes far from the Bragg frequency do not undergo Bragg scattering and they just experience an increase of the mean losses. For the Bragg mode, the threshold increases more than that of the other modes because Bragg scattering introduces a feedback which interferences destructively with the one provided by the Fabry-Perot cavity. As $\delta L$ is increased up to $\delta L \simeq 1.35$ all modes go below threshold and laser death is observed [Fig.(c) and (d)]. As the mean losses $\delta L$ are further increased up to $\delta L \simeq 1.487$, laser revival is observed [see Fig.3(e) and (f)]. The revival is associated to the coalescence of  the two longitudinal modes of the cavity nearest to the Bragg frequency (labelled by -1 and 1 in the figures), and their subsequent splitting along the imaginary axis; see the arrows in Fig.3, which show the motion in the complex plane $q$ of the three poles 0 and $ \pm 1$. The coalescence and subsequent splitting of the poles resemble the EP scenario of Ref.\cite{r1}, however here the poles are resonance states and do not belong to the point spectrum of the Hamiltonian like in Ref.\cite{r1}. Most importantly,  as noticed in the previous case, i.e. for $R=0$, loss-induced lasing is not at all  associated with coalescence of poles in complex plane, i.e. loss-induced lasing is not an effect peculiar to EPs.  

 \par
 In conclusion, loss-induced lasing is not a new result in laser theory and it is not necessarily related to the physics of exceptional points. The interest on such an unusual behavior found in the recent literature seems over-emphasized. The experiment of Ref.\cite{r3} surely provides a nice realization in an integrated micro/nano-scale photonic system of some exotic properties of  non-Hermitian degeneracies, however it does not actually introduce any surprising physical effect into what we know in laser theory.

\end{document}